\documentclass[reviewcopy]{PHYEAUTH}
\usepackage{graphicx}          % include pictures
\usepackage{amsmath}           % math-enviroment
\usepackage{amssymb}           % math-fonts

\begin{document}

\begin{frontmatter}
%\title{Coherent transport of interacting one-dimensional electrons\\ through a single scatterer}
\title{Tunneling of interacting one-dimensional electrons through a single
scatterer: Luttinger liquid behavior in the Hartree-Fock model}

\author{Andrej Gendiar},
\author{Martin Mo\v{s}ko},
\author{Pavel Vagner},
\author{Radoslav N\'{e}meth},

\address{Institute of Electrical Engineering, Slovak Academy of
Sciences, SK-841~04~Bratislava, Slovakia}

\begin{abstract}
We study tunneling of weakly-interacting spinless electrons
at zero temperature through a single  $\delta$ barrier in one-dimensional
wires and rings of finite lengths. Our numerical calculations are based
on the self-consistent Hartree-Fock approximation, nevertheless, our
results exhibit features known from correlated many-body models.
In particular, the transmission in a wire of length $L$ at the Fermi
level is proportional to $L^{-2\alpha}$ with the universal power $\alpha$
(depending on the electron-electron interaction only, not on the
strength of the $\delta$ barrier). Similarly, the persistent current
in a ring of the circumference $L$ obeys the rule $I\propto L^{-1-\alpha}$
known from the Luttinger liquid and Hubbard models. We show
that the transmission at the Fermi level in the wire is related to the
persistent current in the ring at the magnetic flux $h/4e$.
\end{abstract}

\end{frontmatter}

%%%%%%%%%%%%%%%%%%%%%%%%%%%%%%%%%%%%%%%%%%%%%%%
% main text
% \appendix

% \section{}
% \label{}
%%%%%%%%%%%%%%%%%%%%%%%%%%%%%%%%%%%%%%%%%%%%%%%

It is known that a one-dimensional (1D) wire which is free of
impurities and biased by macroscopic contacts yields the quantized 
conductance.  This effect can be derived within a simple model
of non-interacting electrons if we assume a negligible electron
back-scattering. Placing a single impurity (scatterer) into the wire,
the conductance is not quantized any more. It is because of the
back-scattering of the electrons from the scatterer.
The wire conductance (given by the Landauer formula) is proportional
to the electron transmission at the Fermi level~\cite{Datta-95}.

When the electron-electron (e-e) interaction is considered,
the conductance of an infinitely long wire with a single
scatterer decreases as $T^{-2\alpha}$ for ($T\rightarrow0$).
This behavior is known from the Luttinger liquid
model~\cite{Kane-92} where the power $\alpha$ depends on the
e-e interaction only. Assuming a repulsive interaction
($\alpha>0$), the scatterer becomes impenetrable
at zero temperature regardless the strength of the scatterer.

Matveev et al. \cite{Matveev-93} studied the Landauer conductance
of the interacting 1D electrons through a $\delta$ barrier in a
wire with contacts. They analyzed the effect of the Hartree-Fock
potential on the tunneling transmission assuming a weak e-e
interaction. They derived the transmission using the renormalization
group (RG) approach and confirmed the universal power law $T^{-2\alpha}$.
It is believed that this approach goes beyond the Hartree-Fock
approximation.

In this paper we consider the non-Luttinger liquid model of the
same type as analyzed by Matveev et al.~\cite{Matveev-93}.
However, we do not use the RG theory. We apply the self-consistent
Hartree-Fock solution by means of numerical calculations
instead. We calculate the transmission probability at zero
temperature which is related to the Landauer conductance. A good
agreement with the theory of Matveev et al. is found. In particular,
we simulate dependence of the conductance on the wire length ($L$)
for various $\delta$ barriers and we reproduce the universal power
law $\propto L^{-2\alpha}$ that becomes asymptotic for large
wire lengths and/or strong  $\delta$ barriers.

We also consider a mesoscopic ring threaded by the magnetic flux.
As a consequence, the persistent current ($I$) arises~\cite{Imry}.
We study the persistent current of interacting spinless electrons
with a single $\delta$ barrier at zero temperature.
We show how the transmission can be extracted from the persistent
current in the 1D ring where $I\propto L^{-\alpha-1}$ and compare
it with the transmission obtained from the 1D wire.

First, consider 1D wire of length $L$ with interacting
spinless electrons. Both wire ends are connected to contacts
and the single $\delta$ barrier is localized in the center of
the wire. The single-electron wave functions $\psi_k(x)$, where $k$
is the electron wave vector, are described by the Hartree-Fock
equation $H\psi_k(x)=\varepsilon_k\psi_k(x)$. The Hamiltonian
has the form
\begin{equation} \label{Schr}
H = -\frac{\hbar^2}{2m}\ \frac{d^2}{dx^2} + \gamma\delta(x-L/2)
+ U_H(x) + U_F(k,x),
\end{equation}
where $\gamma\delta(x-L/2)$ represents the localized scatterer.
We apply the boundary conditions
\begin{eqnarray} \label{BCondgt}
\nonumber
\psi_k(x=0)&=&e^{ikx}+r_k e^{-ikx}, \\
\nonumber
\psi_k(x=L)&=&t_k e^{ikx}, \\
\label{BCondls}
\nonumber
\psi_{-k}(x = 0)&=&t'_k e^{-ikx}, \\
\psi_{-k}(x = L)&=&e^{-ikx}+r'_k e^{ikx}
\end{eqnarray}
with $r_k$ and $t_k$ being the reflection and transmission amplitudes,
respectively and $k >0$. Knowing the transmission, we can easily
evaluate the Landauer conductance $(e^2/h) \left| t_{k_F} \right|^2$.

The Hartree potential induced by the $\delta$ barrier reads
\begin{eqnarray}
\label{U_H} \nonumber
U_H(x)&=&
\int \limits _{0} ^{L} dx' \ V(x-x') \\
&\times& \int \limits_{-k_F}
^{k_F} \frac{dk'}{2\pi} \left[\left| \psi_{k'}(x') \right|^2 -
\left| \psi^0_{k'}(x') \right|^2 \right]
\end{eqnarray}
and the Fock non-local exchange potential is written as
\begin{eqnarray} \label{U_F1} \nonumber
U_F(k,x) &=& -\frac{1}{\psi_k(x)} \int \limits _{0} ^{L} \!\!
dx' \ V(x-x') \\
&\times& \int \limits _{-k_F} ^{k_F} \!\! \frac{dk'}{2\pi} 
\psi_k(x') \psi^{*}_{k'}(x') \, \psi_{k'}(x)
\end{eqnarray}
with $V(x-x')$ being the e-e interaction. Essentially the same
one-dimensional model was considered by Matveev et al.~\cite{Matveev-93}.

The wire is connected to large contacts via adiabatically tapered
non-reflecting connectors~\cite{Datta-95}.
In the case of interacting electrons without presence of the
scatterer ($\gamma = 0$), we assume that there is no back-scattering
at the wire ends due to the adiabatically tapered connectors.
Then, the solution of Eq.~\eqref{Schr} is described by the
free wave, $\psi^0_{k}(x) = e^{ikx}$, having the eigenenergy
\begin{equation} \label{e_k}
\varepsilon_k = \hbar^2k^2/2m + U_F^0 (k)
\end{equation}
with $U_F^0 (k)\equiv U_F[\psi_{k}(x)= \psi^0_{k}(x)]$ being the
nonzero Fock shift. Note that this solution is valid if we implicitly
assume that the Fock interaction is present also in the contacts.
If the energy in Eq.~\eqref {e_k} holds inside the wire and
we turn off the Fock shift to zero outside the wire, we obtain at
each wire end the potential drop $U_F^0 (k)$. This would cause
back-scattering at both wire ends and the solutions $e^{ikx}$ and
$e^{-ikx}$ would be no longer valid (in contrast to the
ballistic conductance of clean wires~\cite{Datta-95}).

If the barrier $\gamma\delta(x-L/2)$ is positioned in the wire,
this $\delta$ barrier induces Friedel oscillations of
the Hartree-Fock potential.
The Friedel oscillations penetrate through the wire ends into the
contacts, where they decay fast due to the higher dimensionality
of the contacts and decoherence. To mimic this decay within our
model, we sharply turn off the oscillations to zero at both wire
ends keeping $U_H=0$ and $U_F=U_F^0 (k)$ outside the wire. Such a
constant potential emulates the non-reflecting connectors and
justifies the above boundary conditions.

Our numerical results are carried out for the GaAs wire with
the corresponding effective electron mass $m=0.067$~$m_0$,
the electron density $n=5 \times 10^7$ m$^{-1}$, and the
short range e-e interaction
\begin{equation} \label{VeeExp}
V(x - x') = V_0 \,  e^{- \left| x - x' \right|/d}.
\end{equation}
We use the short range e-e interaction (\ref{VeeExp}) because
of comparison with the RG theory~\cite{Matveev-93}
where the e-e interaction is assumed to be finite.
Physical meaning of the finite range is the screening.

The asymptotic formula for the transmission probability at the
Fermi level derived for weak e-e interactions~\cite{Matveev-93} reads
\begin{equation} \label{t-Glazman-Fermi}
\left| {t}_{k_F} \right|^2 = \frac{\left| \tilde{t}_{k_F}
\right|^2 \, (d/L)^{2\alpha}}
     {{ \left| \tilde{r}_{k_F} \right|^2 +
      \left| \tilde{t}_{k_F} \right|^2 \,
      (d/L)^{2\alpha} }
}\simeq \frac{\left| \tilde{t}_{k_F} \right|^2}{|
\tilde{r}_{k_{F}}|^2}\left(d/L\right)^{2\alpha},
\end{equation}
where $d$ is the range of the e-e interaction $V(x-x')$ and
$\tilde{t}_k$ and $\tilde{r}_k$ describes the transmission
and the reflection amplitudes of the bare $\delta$
barrier~\cite{Matveev-93}, respectively. The right hand side of
Eq.~(\ref{t-Glazman-Fermi}) remains valid for small
$\tilde{t}_{k_F}$ and/or large $L$. For weak e-e interaction
($\alpha \ll 1$), the power $\alpha$ reads~\cite{Matveev-93}
\begin{equation} \label{alpha-Glazman}
\alpha = \frac{V(0)-V(2k_F)} {2\pi \hbar v_F},
\end{equation}
where $V(q)$ is the Fourier transform of the e-e interaction
$V(x-x')$. We evaluate $\alpha$ for our e-e
interaction~\eqref{VeeExp}, for which $V(q) = 2V_0 d/(1+q^2d^2)$.

The bare amplitudes are $\tilde{t}_k = k/(k+i\zeta)$ and
$\tilde{r}_k = -i\zeta/(k+i\zeta)$, where $\zeta = \gamma
m/\hbar^2$. As $k_F$ and $m$ remain fixed, we parametrize the
bare $\delta$ barrier by its transmission coefficient
$\left| \tilde{t}_{k_F} \right|^2$ in the following.

Second, consider a 1D wire for which the Hartree-Fock equation
reads
\begin{multline}\label{Eqs-Schroding-2}
\bigg[
\frac{\hbar^2}{2m}\left(-i\frac{\partial}{\partial x}+
\frac{2\pi e}{Lh}\phi\right)^2 +
\gamma\delta(x-L/2)
\\
+U_H(x)+U_F(j,x) \bigg] \psi_j(x) = \varepsilon_j\psi_j(x)
\end{multline}
satisfying the boundary condition $\psi_j(x+L)=\psi_j(x)$.
The persistent current is given as
$I= -\frac{\partial}{\partial \phi}E(\phi)$, where
\begin{equation} \label{Eqs-Vseob-Prud-b}
  E=\sum \limits _j
\left[\varepsilon_j-\frac{1}{2}\left\langle\psi_j
    \left|U_H(x)+U_F(j,x)\right|\psi_j\right\rangle
\right].
\end{equation}

In case of the non-interacting electrons, the persistent
current can be evaluated~\cite{Gogolin} as follows
\begin{equation} \label{I-nonint-approx}
I = (ev_F/2 L) |\tilde{t}_{\varepsilon_{F}}| \sin(2\pi e \phi/h),
\end{equation}
with the transmission amplitude of the scatterer at the Fermi
energy $|\tilde{t}_{\varepsilon_{F}}| \ll 1$ and the Fermi
velocity $v_F$. Assume $L\to\infty$, the repulsive e-e
interaction $\alpha>0$, and $\phi=h/4e$. Then, replacing
$|\tilde{t}_{\varepsilon_{F}}|$ by $|t_{k_{F}}|$ and
inserting Eq.~\eqref{t-Glazman-Fermi} into Eq.~\eqref{I-nonint-approx},
we obtain~\cite{Gogolin}
\begin{equation} \label{I-Luttinger}
I=\frac{ev_F}{2L}|t_{k_{F}}|=\frac{ev_F}{2L}
\frac{|\tilde{t}_{k_F}|}{|\tilde{r}_{k_F}|}\left(\frac{d}{L}\right)^\alpha
\propto L^{-\alpha-1}.
\end{equation}

We solve Eqs.~\eqref{Schr} and ~\eqref{Eqs-Schroding-2} by means
of the self-consistent iterative procedure and follow approximation
in Ref.~\cite{Cohen-97} in order to decrease computational time
and memory. One can further simplify Eq.~\eqref{U_F1} to
\begin{equation} \label{U_F2}
U_F(x) \simeq -\int \limits _{0} ^{L} dx' V(x-x') 
\int
\limits _{-k_F} ^{k_F} \frac{dk'}{2\pi} \text{Re}
[\psi^{*}_{k'}(x')  \psi_{k'}(x)]
\end{equation}
noticing that $\int _{-k_F} ^{k_F} dk' \psi^{*}_{k'}(x') \,
\psi_{k'}(x) \simeq 2\pi \delta(x-x')$. The Fock
potential~\eqref{U_F2} becomes local and independent on $k$.
This allows us to simulate longer wires than in the case
of the non-local Hartree-Fock model, cf.~Eq.~\eqref{U_F1}.
We also present the results without this simplification
but for a substantially shorter wire lengths.

\begin{figure} [tb]
\begin{center}
\includegraphics[clip,width=0.8\columnwidth]{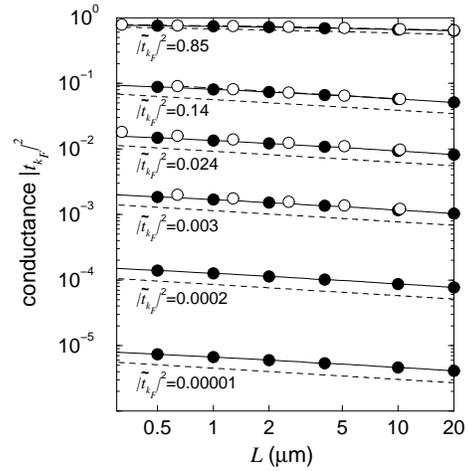}
\end{center}
\vspace{-0.2cm}
\caption{Conductance in unit of the transmission probability at the
Fermi level vs length for various $\delta$ barriers
$\left| \tilde{t}_{k_F}\right|^2$ in a log-log scale. The dashed curves
show the data corresponding to the formula~\eqref{t-Glazman-Fermi}.
The filled circles connected by the full lines are our self-consistent
Hartree-Fock data for the wire; the curves approach the same asymptotic
power law as the dashed lines. The open circles are extracted from the
persistent current using Eq.~\eqref{I-Luttinger}.} \label{Fig:1}
\end{figure}

Figure \ref{Fig:1} shows the transmission probability $\left|
{t}_{k_{F}} \right|^2$ for the wire and the ring versus the length
$L$ for various $\delta$ barriers. The result of the
RG~\eqref{t-Glazman-Fermi} is presented by the dashed lines. For
strong $\delta$ barriers the dashed lines follow the asymptotic
power law $\left| t_{k_F} \right|^2 \propto L^{-2\alpha}$ as
manifested by the linear decay with slope $-2\alpha$ in the
log-log scale. Our Hartree-Fock curves (filled and open circles)
show slightly higher transmission following the same slope $-2\alpha$.
Note that for $\left| \tilde{t}_{k_F} \right|^2$ small enough
and substantially longer wires and rings, all the Hartree-Fock
curves decay with the same slope as the dashes curves,
independently on the strength of the $\delta$ barrier.

\begin{figure}[tb]
\begin{center}
\includegraphics[clip,width=0.95\columnwidth]{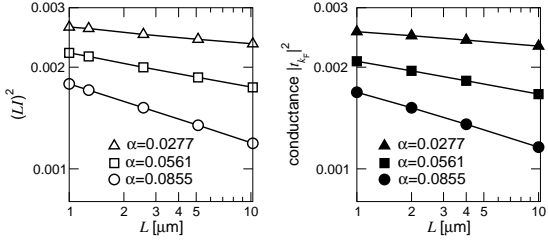}
\end{center}
\vspace{-0.1cm}
\caption{A log-log plot of the persistent current vs ring circumference
(left) and the conductance (transmission) vs the wire length for
different $\alpha$.} \label{Fig:2}
\end{figure}

In Fig.~\ref{Fig:2} we present the squared persistent current
(which has the meaning of transmission, cf.~Eq.~\eqref{I-Luttinger})
and the conductance as functions of
the length $L$ for both the ring (left) and the wire (right). The
data shown were calculated for the bear $\delta$ barrier
$\left| \tilde{t}_{k_F} \right|^2=0.003$ and for various e-e
interactions. As seen, both data for the wire and the ring follow
the same behavior having the slope $-2\alpha$. We have chosen
the e-e interactions $\alpha=0.0277$, 0.0561, and 0.0855 that
correspond to $V_0=11$~meV, 22.3~meV, and 34~meV, respectively,
and for the common range of e-e interaction $d=3$~nm.

\begin{figure}[tb]
\begin{center}
\includegraphics[clip,width=0.9\columnwidth]{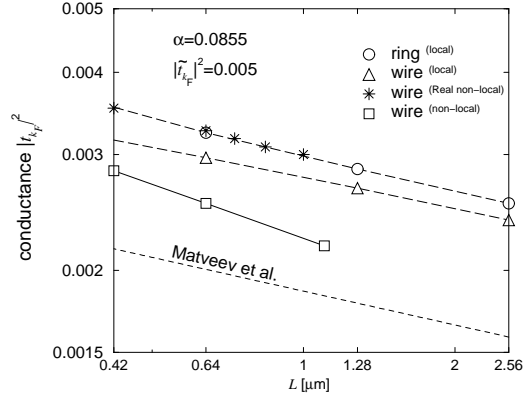}
\end{center}
\vspace{-0.3cm}
\caption{Conductance vs. length of the system for the wires and the
rings within local and non-local Fock approximations for $\alpha=0.0855$ and
$\left| \tilde{t}_{k_F} \right|^2=0.005$. The dashed line shows
the analytical formula~\eqref{t-Glazman-Fermi} of Matveev et al.}
\label{Fig:3}
\end{figure}

The presented results were calculated
within the local Fock term (Eq.~\eqref{U_F2}).
When the non-local Fock term is considered, the resulting
effective potential gains a nonzero imaginary part which
changes the transmission properties of the system
studied. In Fig.~\ref{Fig:3} we depicted the conductance at
short lengths for the wire (triangles) and the ring (circles)
within the local approximation~\cite{Cohen-97}, the wire for
the non-local Fock term (squares) and the non-local Fock term
for which the imaginary part is omitted (stars).

If the Fock term is completely omitted, we obtain different
results for the conductance (or the persistent current). It
is because the perfect reflection is resulted instead (in the
case $L\to\infty$). However, the power-law behavior is expected
to persist if the density functional theory is applied.

Note that we have not yet reached the asymptotic regime
because the conductance in Fig.~\ref{Fig:3} is not linear
in the log-log scale. Longer wire lengths are required when
the non-local Fock term is involved in our calculations.
Unfortunately, the memory limitations (4~GB) do not
enable us to obtain reliable data for longer systems.

In conclusion, we have calculated the Landauer conductance and
the persistent current of the weakly-interacting spinless electrons
in the 1D wires and rings with a single $\delta$ barrier. We have
used the self-consistent
Hartree-Fock approximation at zero temperature. We have found
the universal power law $\left| t_{k_F} \right|^2 \propto
(d/L)^{2\alpha}$ known from the Luttinger-liquid model~\cite{Kane-92}
and the RG models~\cite{Matveev-93,Meden-02}. We conclude
that the universal power law is not exclusively the consequence
of correlations; it can be obtained within the Hartree-Fock
approximation. To prove this, longer lengths are needed.
We have also found that essentially the same wire
conductance can be extracted from the persistent current in the
1D rings.

The APVT grant
APVT-51-021602 and the VEGA grant 2/3118/23 are acknowledged.

%%%%%%%%%%%%%%%%%%%%%%%%%%%%%%%%%%%%%%%%%%%%%%%
% bibliography
%
% notes:
% \bibitem{label2}
% If there is a note, it should come last:
% \bibitem{label3} \note
% \end{subbibitems}
%%%%%%%%%%%%%%%%%%%%%%%%%%%%%%%%%%%%%%%%%%%%%%%

\end{document}